\documentclass[10pt,aps,prb,amsmath,amssymb,reprint,letterpaper,nobalancelastpage,final,citeautoscript,floatfix,raggedbottom,longbibliography]{revtex4-1}
\usepackage[usenames,dvipsnames]{color}
\usepackage{graphicx}
\usepackage{microtype}
\usepackage[bookmarks=false,colorlinks]{hyperref}
\usepackage{xfrac}
\hypersetup{linkcolor=RubineRed,citecolor=RoyalBlue,filecolor=Mulberry,urlcolor=RoyalBlue}

\newcommand{\revision}[1]{\textcolor{black}{#1}}

\begin{document}

\title{
Symbolic regression in materials science}

\author{Yiqun Wang}
\thanks{These two authors contributed equally to this work.}
\affiliation{Department of Materials Science and Engineering, Northwestern University, Illinois 60208, USA}

\author{Nicholas Wagner}
\thanks{These two authors contributed equally to this work.}
\affiliation{Department of Materials Science and Engineering, Northwestern University, Illinois 60208, USA}

\author{James M.\ Rondinelli}
\affiliation{Department of Materials Science and Engineering, Northwestern University, Illinois 60208, USA}

\begin{abstract}
\revision{We showcase the potential of} symbolic regression as an analytic method for use in 
materials research. 
First, we briefly describe the current state-of-the-art method, 
genetic programming-based symbolic regression (GPSR), and 
recent advances in symbolic regression techniques. 
Next, we discuss industrial applications of symbolic regression and 
its potential applications in materials science.
\revision{We then present two GPSR use-cases: formulating a transformation kinetics law 
and showing the learning scheme discovers the well-known Johnson-Mehl-Avrami-Kolmogorov 
(JMAK) form, and learning the Landau free energy functional form for the displacive tilt transition in perovskite LaNiO$_3$.}
Finally, we propose that symbolic regression techniques should be considered by 
materials scientists as an alternative to other machine-learning-based regression
models for learning from data.
\end{abstract}

\maketitle

\section{Motivation} \label{sec:1}

\subsection{Era of big data in materials science}
Modern scientists perpetuate 
the scientific process embodied by the works of 
Tyco Brahe, Johannes Kepler, and Isaac Newton in the heliocentric revolution. 
Brahe was the observationalist. He took extensive, precise measurements of the 
position of planets over time. 
Kepler was the phenomenologist. From Brahe’s measurements, he derived concise 
analytical expressions that describe the motion of the solar system in a succinct manner. 
Last, Newton was the theorist.
\revision{He realized the mechanism behind the apple falling from the tree is the same as that underlying planets traveling around the sun, which could be formulated into a universal law (Newtonian gravitational law).}
All three scientific modalities are vital in making scientific discoveries: 
data acquisition (Brahe), data analysis (Kepler), and derivation from 
first-principles (Newton). 

With recent advances in computer science, theoretical modelling, and experimental instrumentation, 
materials scientists have in many ways created a ``mechanical Brahe'' and 
marched into a new era of big data. 
Datasets of materials information, obtained from advanced characterization techniques\cite{deelman2017, lupini2018, ren2017}, 
combinatorial experiments\cite{Stein2019, Alberi2019, Green2017}, 
high-throughput first-principles simulations\cite{Ye2018, Tanaka2018}, 
literature mining\cite{Kim2017, Krallinger2017}, and other techniques, 
are created at a faster rate every day with less and less human labor. 
All of this data enables new opportunities to construct novel 
laws of phenomenological behavior for systems that previously lacked them. 
Inspired by the Materials Genome Initiative (MGI) \cite{Government2014}, 
the materials community is working collaboratively towards making 
digital materials data accessible to others. 
\revision{Multiple materials databases such as \href{https://materialsproject.org}{Materials Project}\cite{Jain2013}, \href{http://oqmd.org}{OQMD}\cite{Saal2013MaterialsOQMD}, \href{http://www.aflowlib.org/}{AFLOWLIB}\cite{Curtarolo2012}, \href{https://omdb.mathub.io/}{OMDB}\cite{Borysov2017}, \href{http://www.aiida.net/}{AiiDA}\cite{Pizzi}, \href{https://citrination.com}{Citrination} and \href{https://nomad-coe.eu/}{NOMAD},
provide public access to millions of materials data points.}
Accessibility to an immense amount of materials data paves way for the next 
step of ``automating Kepler'' in the discovery of governing laws in materials
processing-structure-properties-performance relationships, 
which could advance materials discovery, development, and technology innovation.

Since one of the fundamental research objectives of materials science and engineering 
is to deliver new materials with optimal performance under specified constraints, 
it is essential to understand how and which features govern the functionality.
In other words, which 
degrees-of-freedom (or parameters)  and their corresponding intrinsic 
relationships (or dependencies) to the material properties should  be optimized.
However, the multi-scale nature of materials science\cite{Alberi2019}, 
e.g.\ from atomic-scale crystal structure to complex mesoscale domain structures and 
bulk mechanical properties or from femotosecond laser probes to 
hour-long recrystallization reactions, makes it particularly challenging 
to study many hierarchical relationships of different materials families.
Given such a high-dimensional parameter space (e.g.\ chemical composition, 
crystal structure, external conditions, etc.), materials scientists often 
explore a finite subspace of all the factors that govern materials properties 
and performance.
In addition, the available data is typically sparsely distributed. 
Although, access to a large materials database relieves, in part, the 
limited-data problem, there is an urgent need for a robust 
data-processing protocol to help discern governing laws in materials science 
and  to deliver designer materials and synthesis/processing procedures.

\subsection{An alternative to machine-learning methods}
Much of the burgeoning field of materials informatics focuses on 
the aforementioned challenges. 
Machine learning (ML) models are currently the tools of choice 
for uncovering these physical laws. 
Although they have shown some promising performance in predicting 
materials properties \cite{Zhuo2018}, typical parameterized machine 
learning models are not conducive to the next stage of generalizing 
across domains---the ultimate goal of ``automating Newton.'' 

\begin{figure}[t]
  \centering
 \includegraphics[width=0.98\linewidth]{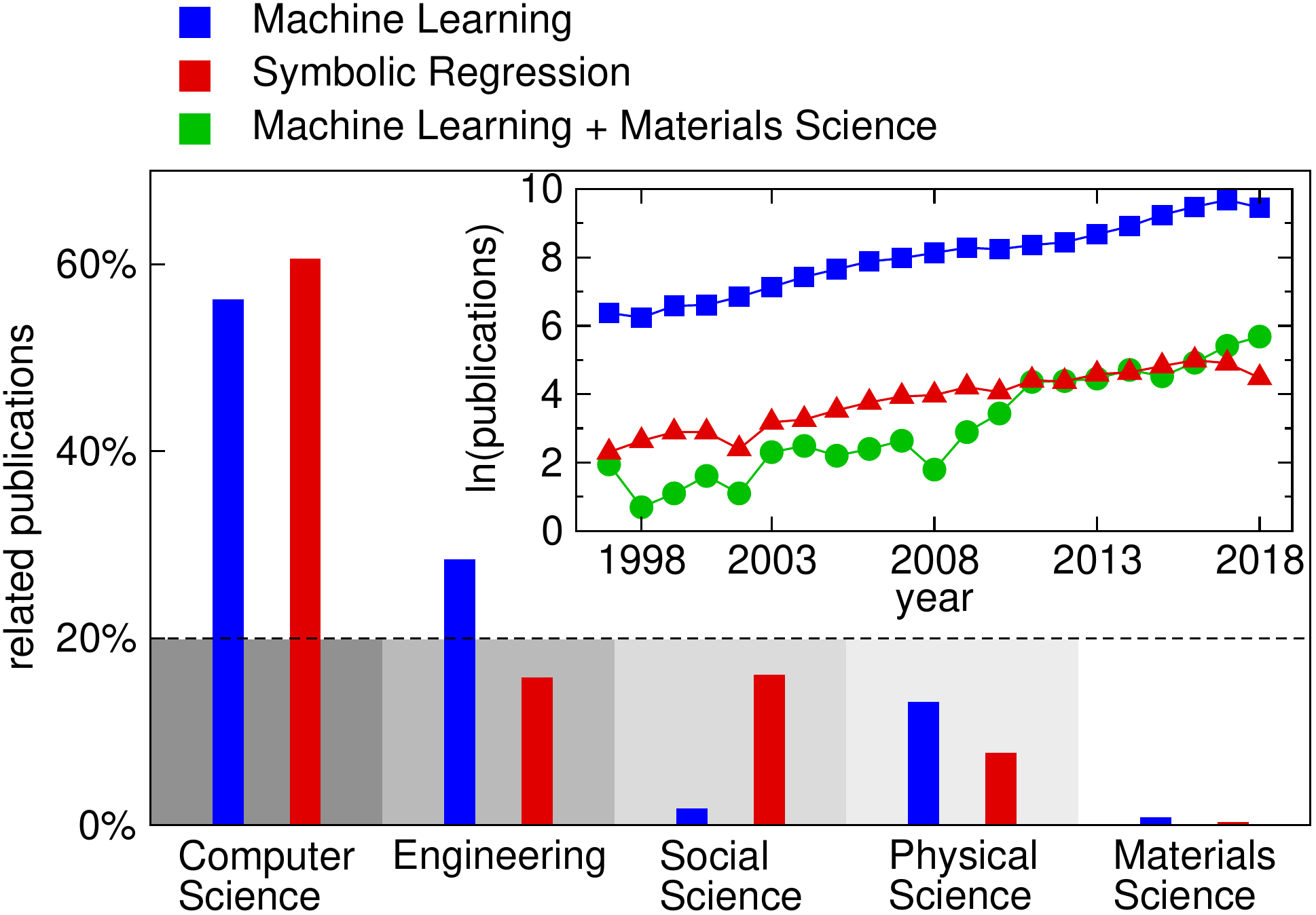}
  \caption{Relative contribution from different research domains to scientific journals related to machine learning and symbolic regression. Shaded panels  indicate a  20\% level of the research domain, emphasizing an opportunity in materials science. (inset) The trend in number of related publications on a natural logarithmic scale (ordinate) related to machine learning, machine learning and materials science, and symbolic regression, with respect to time.}
  \label{fig:publication}
\end{figure}

It is important to note that Newton's challenge was somewhat 
made easier, because Kepler's laws were parsimonious yet predictive. 
In a modern context, ML models can be predictive but their descriptions 
are  often too verbose (e.g.\ deep-learning models with 
thousands of parameters) or mathematically restrictive (e.g.\ 
assuming the target variable is a linear combination of input features).
Such black-box models have become more prevalent in modern 
materials science research; however, the interpretability 
of such models have always been a problem.
Although there is a large body of work on data visualization and model 
understanding to address these issues, those subjects will be out of the scope of this perspective (see Ref.\ \onlinecite{Hall2018} for a review).
In this prospective paper, we focus on an alternative to
machine-learning models: symbolic regression. 
Symbolic regression simultaneously searches for the optimal form of 
a function and set of parameters to the given problem, and is a powerful 
regression technique when little if any a-priori knowledge 
of the data structure/distribution is available.
\autoref{fig:publication} shows the relative popularity of machine 
learning and symbolic regression in different research domains. 
We use data from the ``Web of Science Core Collection'' database 
in this analysis \cite{webofscience}. 
Among all publications whose topics are related to machine learning or 
symbolic regression, over 50\% of the contributions come from the 
computer science research community, while multidisciplinary engineering 
is second. 
Social science and physical science each makes less than 20\% of 
the contribution to the total number of publications. 
These two techniques are not so popular in materials science research, 
as the relative contribution is almost negligible compared to other 
research domains. 

It is not surprising to see a dominant contribution from 
computer science in both the machine learning and symbolic 
regression communities, since it is where these techniques were 
born. 
\revision{It is interesting to notice that symbolic regression is 
relatively more popular than machine learning in social science research.}
%
One possible reason for this trend is that social science problems 
typically do not have a (known) physically motivated 
governing equation as in many physical sciences, where for example, 
Newtonian equations-of-motion, Schrodinger equation, etc.\ can 
be written formally. 
Symbolic regression arises naturally as a problem solver 
since it has the potential to find an appropriate functional 
form from social science data sets, e.g.\ questionnaire results, 
behavior patterns, etc. 

\revision{We also report the trend in the number of publications (in 
a natural logarithm scale) in the following research domains 
[\autoref{fig:publication}(inset)]: 
machine learning, 
application of machine learning in materials science,
and symbolic regression. 
All three domains exhibit a rapid (almost exponential) growth rate, 
whereas the number of machine-learning-related publications is orders of magnitude larger than the other two.
The trend of symbolic regression applications in materials science 
is not shown here since the base number is too small; nonetheless, it 
also reveals a potential previously underappreciated research domain. 
For materials science problems, one is often also presented with 
the problem of unknown relationships among many variables.
%
Symbolic regression presents an opportunity then to help in the 
formulation of structure-property relationships derived from 
these variables.} 

\revision{In this prospective paper, we encourage materials scientists and 
engineers to utilize symbolic regression techniques in solving their 
domain problems.
To facilitate a better understanding of the utility and 
application of symbolic regression, we next 
introduce the genetic programming-based symbolic regression (GPSR) 
method and describe current research frontiers in symbolic regression. 
Next, we 
discuss several industrial applications of symbolic regression and 
propose potential uses in materials science. 
In addition, we present how GPSR can learn the 
Johnson-Mehl-Avrami-Kolmogorov (or Avrami) equation to describe 
recrystallization kinetics, as well as the Landau free energy expansion describing the structural phase transition in LaNiO$_3$.
Last, we conclude with some open challenges in materials research 
that may benefit from symbolic-regression methods.}

\section{Symbolic regression and current state-of-the-art methods}\label{sec:2}
 
\subsection{Genetic programming-based symbolic regression (GPSR)}

Symbolic regression is a method of finding a suitable mathematical 
model to describe observed data \cite{Augusto2000}. 
In conventional regression techniques, one optimizes parameters for 
a particular model provided as a starting point to the algorithm. 
For instance, a linear regression model is based on the 
assumption that the relationship of the dependent variables and 
regressor is linear\cite{Seber2003}; 
an artificial neural network (ANN) is a nonlinear model which 
relies on a predefined network infrastructure such as 
neuron connections and activation function 
(e.g.\ sigmoid, softmax function). 
In symbolic regression, however, no such a-priori assumptions 
on the specific form of the function is required. 
Instead, one provides a 
mathematical expression space containing candidate function building 
blocks, e.g.\ mathematical operators, state variables, constants, 
analytic functions, and then symbolic regression searches through the 
space spanned by these primitive building blocks to find the most 
appropriate solution. 
In other words, both model structures and model parameters are 
optimized in symbolic regression. 
Since there is no need for a predefined function form, optimization 
algorithms used in symbolic regression are different from conventional 
analytical/numerical optimization methods (e.g.\ conjugate gradient, 
Newton-Raphson method). 
In this section, we briefly introduce one of the most prevalent methods used in symbolic regression by means of genetic programming.

\begin{figure}[t]
  \centering
 \includegraphics[width=0.9\linewidth]{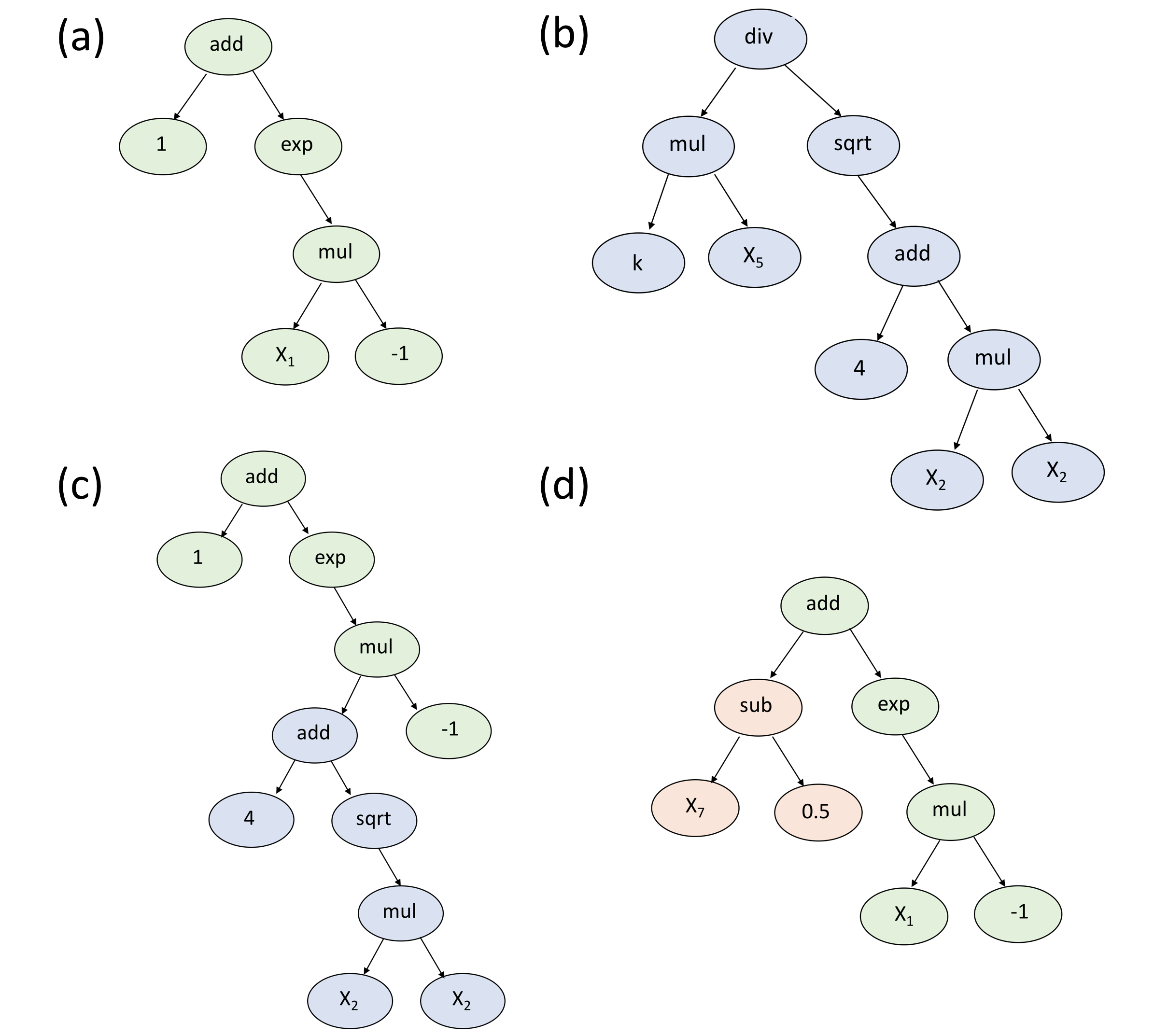}
  \caption{Tree-structure chromosome representation of computer programs in genetic programming. (a) parent1 ($1+\exp(-x_1)$); (b) parent2 (${kx_5}/{\sqrt{x_2^2+4}}$); (c) child of genetic crossover operation ($1+\exp(-\sqrt{x_2^2+4})$); and (d) child of subtree mutation operation ($x_7-0.5+\exp(-x_1)$).}
  \label{fig:GPSR}
\end{figure}

Genetic programming (GP) was developed by J.R.\ Koza \cite{Koza1994} as 
a specific implementation of genetic algorithms (GA) \cite{Forrest1993}, 
which are often utilized in the materials community for atomic structure prediction\cite{Meredig2013, Chua2010, Mohn2011}. 
The idea is to evolve the solution of a given problem following Darwin's theory of evolution and 
to find the fittest solution after a number of generations.
Instead of using strings of binary digits to represent chromosomes as in GA, 
solutions in GP are represented as tree-structured chromosomes with nodes and terminals. 
\autoref{fig:GPSR}a shows a chromosome example of the 
mathematical function $1+\exp(-x_1)$. 
The tree consists of a set of interior nodes with mathematical operations 
($+$, $\times$, $\exp$) and terminal nodes with variables 
($x_1$) and constants ($\pm1$). 
A depth-first search can be used to traverse the tree to get the final mathematical 
expression of each individual solution. 
The structure of a chromosome tree is not necessarily binary; its structure 
depends on the number of arguments the mathematical operator takes. 
For demonstration purposes, we only introduce simple operators that are 
either unary or binary. 
Users of GP could include a variety of functions suitable for their target problems. 
A large number of trees will be generated based on specified user settings 
and evaluated throughout the GP process. 
Each tree represents a potential solution of the problem. 
The way new trees are generated from the initial mathematical building blocks is a 
unique feature of GP since it mimics the natural evolution of Earth's ecosystem, i.e.\ 
through artificial sexual recombination and a natural selection process. 

\begin{figure}[t]
  \centering
 \includegraphics[width=0.9\linewidth]{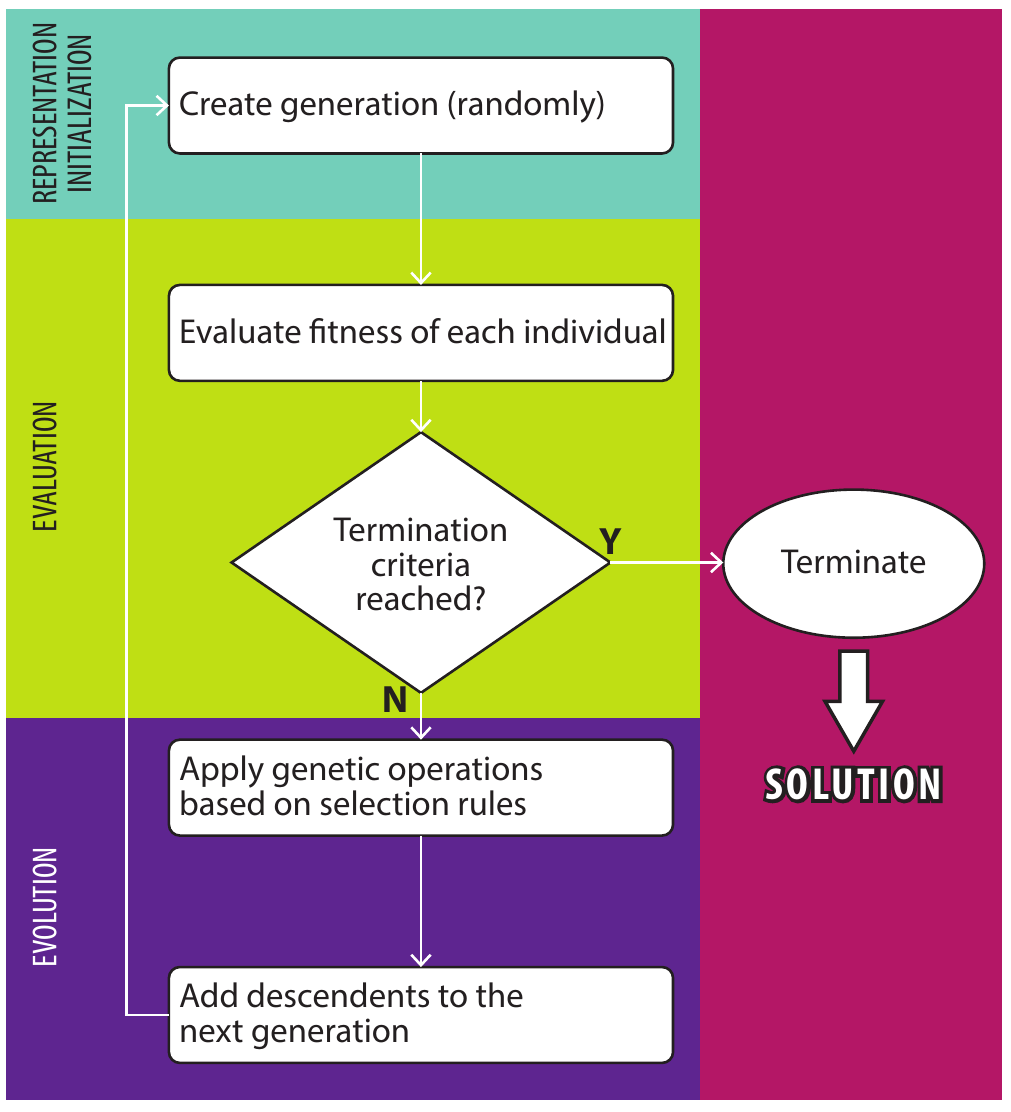} 
  \caption{Genetic programming flowchart depicting the iterative solution-finding process.  	}
  \label{fig:flowchart}
\end{figure}

\autoref{fig:flowchart} illustrates the process by which a solution of 
the symbolic regression problem is obtained using  genetic programming.
The procedure starts with a set of randomly generated initial terminal 
nodes (variables, constants) and functions, forming individual trees with 
different sizes and structures (\autoref{fig:GPSR}a-d). 
These fundamental building blocks come from a user-defined input set. 
This starting population typically has a large variety of tree structures due to the random process, which facilitates further exploration of the variable space and 
reduces the potential risk of being trapped in local minima. 
The initialization process terminates once the number of individuals reaches a user-defined population size, where the natural selection process then 
comes into play. 
The ``fitness'' of each individual solution in the initial population is then 
evaluated by comparing their function output with the true value from the data set.
This fitness value describes how well the program performs in terms of solving the 
problem. 
The common error metrics used include mean squared error (MSE), 
root-mean squared error (RMSE), etc. 
Then GP evolves the current generation by randomly applying genetic 
operations to individuals, e.g.\ crossover and mutation. 
One or more individuals from the current generation will be selected as 
parent(s) based on the fitness score, typically the higher the score, the 
larger probability to be selected for reproduction. 
Such a selection rule agrees with the ``survival-of-the-fittest'' rule since 
good features are more likely to be inherited by the next generation, which is 
the essential step towards the optimal solution.

\begin{table*}[t]
\begin{ruledtabular}
\caption{Comparison of genetic programming-based symbolic regression (GPSR)
and its alternative methods: 
multiple regression genetic programming (MRGP), 
geometric semantic genetic programming (GSGP), 
Cartesian genetic programming (CGP), 
genetic programming-based relevance vector machine (GP-RVM), 
evolutionary polynomial regression (EPR), and 
fast function extraction (FFX).
\label{tab:GP_alters}}
\centering
\begin{tabular}{lllll}
 & \sffamily Program representation & \sffamily Fitness evaluation & \sffamily Optimization & \sffamily Solution form \\
\hline
GPSR & rooted-tree & individual program & genetic evolution & rooted-tree \\
\hline
MRGP & rooted-tree & subexpressions & genetic evolution + & linear combination \\[-0.5em]
& & & linear regression & of subexpressions \\
\hline
GSGP & rooted-tree & distance in & genetic evolution in  & rooted-tree/ \\[-0.4em]
& & semantic space & semantic space & semantic vector \\
\hline
CGP & acyclic graph & individual program & genetic evolution & 2D grid of nodes \\
\hline
GP-RVM & rooted-tree & group of GP & genetic evolution + & linear combination of \\
& & individual & RVM &  GP individuals \\
\hline
EPR & vector of integers & individual program & genetic evolution + & polynomial function \\[-0.4em]
& & & linear regression & \\
\hline
FFX & basis functions & individual program & pathwise regularized & linear combination of \\[-0.4em]
& & & learning & basis functions \\
\end{tabular}
\end{ruledtabular}
\end{table*}

The genetic crossover operation takes two winners of the selection process 
as parents to breed their offspring. 
For instance, the two structures in \autoref{fig:GPSR}a and b are 
taken as parents. 
The crossover operator then randomly takes a subtree from parent (b) 
and substitutes another random subtree in parent (a) with that from (b). 
One possible offspring from the crossover operator is illustrated in
\autoref{fig:GPSR}c. 
Crossover is usually the dominant operation in the recombination process.
\autoref{fig:GPSR}d is an example of an offspring from the mutation operation. 
The mutation operator only takes one parent structure, and randomly 
substitutes a subtree with another randomly generated structure; 
in case (d), the constant 1 is mutated to $(x_7 - 0.5)$. 
Although this operation is more aggressive compared to the crossover 
operation, since it adds randomness to the system, 
it is important to have a finite chance of mutation to introduce new 
variations, e.g.\ new constants and new features, and avoid 
being trapped in local minima. 
The third category of genetic operations is reproduction, which duplicates 
the selected program and directly inserts its offspring to the next 
generation. 
It guarantees that some of the current generation will be preserved by the 
next generation, and partially protects the similarity between two generations.
Detailed definitions and implementations of each genetic operation can vary 
from case to case, but the main features should be the similar to what we 
described here. 

The newborns are then added to the next generation after each genetic 
operation, until the new population size reaches the specified set number. 
Then the new generation goes through the fitness evaluation and natural 
selection process again until the fitness value reaches a certain criteria  
or the maximum number of generations is reached.
After termination of GP, the surviving individuals are expected to be 
highly evolved to adapt to the problem-dependent selection rule. 
More comprehensive descriptions of GP can be found in Koza's original  
paper\cite{Koza1994}.
\subsection{Advances in symbolic regression}
Since Koza introduced the idea of GP in 1992, there have been 
significant efforts made to improve the performance of the original GPSR 
algorithm. 
The major problems to overcome in GPSR include: 
\begin{enumerate}
    \item[($i$)] \emph{Non-deterministic optimization.} It is not guaranteed that the performance of the descendent generations will be better than their parents. 
    \item[($ii$)] \emph{Difficulty in finding the proper constants.} Since the way GP generates constants is random, either in the initial input set or those brought into the population by mutations, there is no effective way to obtain the ideal coefficients as in other numerical regression methods.
    \item[($iii$)]\emph{Limited capability to preserve good components of the equation} due to the fitness evaluation method. The fitness is evaluated based on the complete structure of an individual. Having a good feature in a subbranch does not necessarily lead to better individual performance, thus good equation components may get lost in the next generation. 
\end{enumerate} 
We summarize some of the most popular alternative methods to conventional 
GPSR in \autoref{tab:GP_alters} and discuss their similarities as well as 
the differences in four aspects, namely program representation, fitness 
evaluation, optimization method, and the solution form.
%

Multiple regression genetic programming (MRGP) \cite{Arnaldo2014} improves 
the program evaluation process by performing multiple regression on
subexpressions of the solution functions. 
Instead of evaluating the fitness of each individual solution as a whole, 
MRGP decouples its mathematical expression tree into subtrees. 
The fitness of the solution is evaluated based on the best linear 
combination of these subtree structures. 
A least angle regression (LARS) algorithm is used to solve the linear 
regression problem here. 
Such a fitness evaluation scheme places more emphasis on finding good 
components even though it might only be a partial solution. 
For instance, the individuals that contain a correct form of a subtree 
structure of the correct solution (if known) are more likely to survive the
natural selection process and pass these good features to the descendents.
MRGP essentially decouples the current basis functions to find the best 
solution in an enlarged space at the vicinity of the original GP space. Indeed, this feature may be well-suited for multi-scale materials problems where modeling of systems across different length/time-scale is desired \cite{Moore2014}. While some subexpressions capture relationship among variables within each scale, the final symbolic regression solution assembles models across the scale and returns the multi-scale model.

\revision{
Geometric semantic genetic programming (GSGP) \cite{Castelli2015} evaluates 
the semantic performance of a computer program instead of the syntax 
performance as in conventional GPSR. 
While still using a rooted-tree structure to represent computer programs, 
GSGP focuses on its semantics, 
i.e.\ the behavior of a program. 
For instance, $add(x_1, x_1)$ is equivalent to $mul(2, x_1)$ in semantic 
space, 
but quite different in terms of syntax. 
It is reasonable to care more about the behavior of the program 
than how the function appears.
By representing each program in a high-dimensional semantic space, 
the fitness evaluation is rather straightforward; 
one only needs to measure the distance 
of the program from the target point in that space.
The closer a program is to the target point, 
the better performance it has in solving the given problem.
Interestingly, the offspring of two parent vectors in semantic space 
lies between its parents in the semantic space; 
therefore, the offspring should be at least no worse performing 
than the poor-performing parent.
Optimizing program semantics rather than syntax further frees symbolic 
regression from specific function forms, potentially making SR more efficient \cite{Castelli2015}.
}

Cartesian genetic programming (CGP) \cite{Miller2000} has a more sophisticated design than conventional GP. 
Here, a computer program is represented as a directed acyclic graph, 
which may be visualized as a two-dimensional grid of nodes. 
Each node owns a set of genes that determines the input-output and 
mathematical function that the node performs; the whole set of genes of the 
computer programs form its genotype. 
Decoding the genotype leads to the phenotype, i.e.\ the function form 
of the computer programs. 
The genotype-phenotype mapping is a unique feature of CGP which makes it closer to the real natural process.

GP-RVM \cite{Rad2018} is an alternative GP method that combines Kaizen 
programming and a relevance vector machine (RVM) algorithm to solve 
symbolic regression problems. 
Kaizen programming (KP) is a collaborative version of genetic programming, 
where individuals work together with each other to solve the problem. 
The solution of a Kaizen process is a linear combination of GP individuals, 
and thus the fitness evaluation is based on a group of individual 
partial solutions instead of an individual program as a complete solution.
RVM is a Bayesian kernel method that could extract important basis functions 
from the basis set without the prior knowledge to set a threshold and  
automatically deals with singularity. 
GP-RVM leverages advantages from both evolutionary algorithm and Bayesian 
kernel methods: the former mainly explores the parameter space while the 
latter extracts basis functions to build and solve for the optimal solution 
function within that space.

Evolutionary polynomial regression (EPR) \cite{Giustolisi2009} hybridizes 
the parameter estimation used in conventional numerical regression methods 
with the evolutionary optimization scheme in GPSR. 
EPR first explores the function space using genetic algorithms, then performs 
linear regression (e.g.\ least squares) to optimize the coefficients of each 
mathematical building block. 
Although EPR specifically uses polynomial expansions for the form of the functions, 
the solution is not necessarily a simple polynomial function since the transformed 
variables used in the polynomial expansion could be nonlinear functions of independent 
input variables. 
Such a hybrid method improves the stochastic GPSR method moving it towards 
a more deterministic approach although the computational cost may 
be relatively higher.
In fact, the polynomial form of the expressions could make EPR suitable for materials design or multiobjective optimization purposes. Since the analytical gradient and Hessian of the solution can be evaluated, materials scientists may have more insights regarding the system and know what parameters to tune in order to achieve optimal design.

Fast function extraction (FFX) \cite{McConaghy2011} is an efficient way to find good 
basis functions and solve for the best solution within the space it spans.
%
The first step in FFX is to generate a large number of candidate basis functions built 
from input variables and other predefined variables. 
The evolutionary optimization scheme is not involved in FFX, instead, a pathwise 
regularized learning technique is used to identify the best coefficients and basis 
functions for the solution. 
Then, models obtained from the previous step are assessed based on the validation 
data set as well as their model complexity in order to identify the best solution. 
FFX is more efficient compared to other GP-based methods due to the deterministic 
optimization technique. Materials scientists could first use FFX to see whether the input function/variable basis is sufficient for their research problem, before further investigation using symbolic regression methods (either FFX or other variants).

The performance of some of the recently developed symbolic regression techniques has 
been assessed against popular machine learning methods \cite{Orzechowski2018}, 
and it is reported that symbolic regression performs considerably well 
compared to state of the art ML algorithms with regards to predictive accuracy.
However, the two methods do not simply exist in competition to one another. We also observe a trend of more hybridization between 
conventional ML algorithms and genetic programming in symbolic regression solvers\cite{Icke2013,Krawiec2002,lu2016using,li2019neural}. 
These advances have enabled symbolic regression to be used for solving real-world problems, which we will discuss in the following section.

\section{Applications of symbolic regression} \label{sec:3}

\revision{
Although it seems that equations obtained from first principles (e.g., the Schr\"{o}diner equation) and empirical observations (e.g., the 18-electron rule \cite{tolman197216}) are quite contradictory to each other, we see quite often that they symbiotically work together in solving real-world problems. 
For instance, both \textit{ab initio} and experimental data have been used to develop effective interatomic force fields \cite{van1990force} or exchange-correlation functionals\cite{yanai2004new}. %
In fact, symbolic regression has the potential to serve as the bridge connecting experimental data to first principles. 
Schmidt \textit{et al.} demonstrated that symbolic regression is capable of predicting connections between dynamics of subcomponents of the system and distill natural laws from experimental data \cite{Schmidt2009}.
Moreover, symbolic regression provides researchers with analytic equations, which expectably would have better interpretability 
over the raw data and potentially other black-box models. 
The equations could reveal how the dependent variable (system output) responds to multiple independent variables (system input), as well as the relationships between independent variables of the underlying function. We show this later in Section \ref{sec:use_case}. 
}

\revision{Common motivations underlying the use of GPSR for complex problem solving %
include when the system in question is not effectively modelled by a linear model. Existing multiple linear regression models are much faster and are already easy to interpret. GPSR is best used for systems with complex interactions between observable variables for which the form of which is not known beforehand---a situation common in materials science and engineering.
}

\revision{In addition, a GPSR approach could be useful for design optimization purposes. Although the evolutionary search process is a black box, the final solution is analytical,  which potentially contains important information (e.g., regarding the gradient or Hessian) about relationships between the design variables and objectives. There is also need for multi-objective optimization such as finding the Pareto optimal combination of model performance and complexity in various domains---it is here that the symbolic regression technique has shown to be effective and interpretable \cite{Gout2018}. We next describe some applications of symbolic regression in various science and technology domains.
}

\subsection{Industrial applications}
GPSR has been applied to a wide variety of problems in fields outside of materials 
science and chemistry. 
Most prominently featured in the popular press was work published by 
Schmidt and Lipson in \textit{Science},\cite{Schmidt2009} which showed GPSR could 
discover Hamiltonians and Lagrangians for systems of simple harmonic oscillators and 
double pendulums. 
Reports of using GPSR for real world systems, however, have been published since 
Koza's origination of the idea in the early 1990s and continue today. 
Arkov \textit{et al.}\cite{Arkov2000} used GPSR to identify equations governing gas 
turbine engines under multiple optimization conditions. 
Berardi \textit{et al.}\cite{Berardi2008} used GPSR to find easy to interpret models 
for pipe failures in a UK water distribution system. 
Bongard and Lipson\cite{Bongard2007} applied GPSR  to generate symbolic equations for 
nonlinear coupled dynamical systems in mechanics, ecology, and systems biology. 
The authors also emphasized that their symbolic models are easier to interpret than 
numerical models, which makes understanding more complex systems easier for future 
applications. 
Cai \textit{et al.}\cite{Cai2006} identified correlation equations from experimental 
heat transfer measurements using GPSR with a sparsifying constraint. 
The authors' predicted correlations had lower percentage error than models developed 
graphically and numerically, albeit with more formula complexity than those 
traditional methods. 
Can and Heavey\cite{Can2011} applied GPSR to develop metamodels for predicting 
throughput rates in industrial serial production lines. 
McKay, Willis and Barton developed steady-state models for a vacuum distillation 
column and a chemical reactor\cite{McKay1997}. 

\begin{figure}[t]
  \centering
 \includegraphics[width=0.9\linewidth]{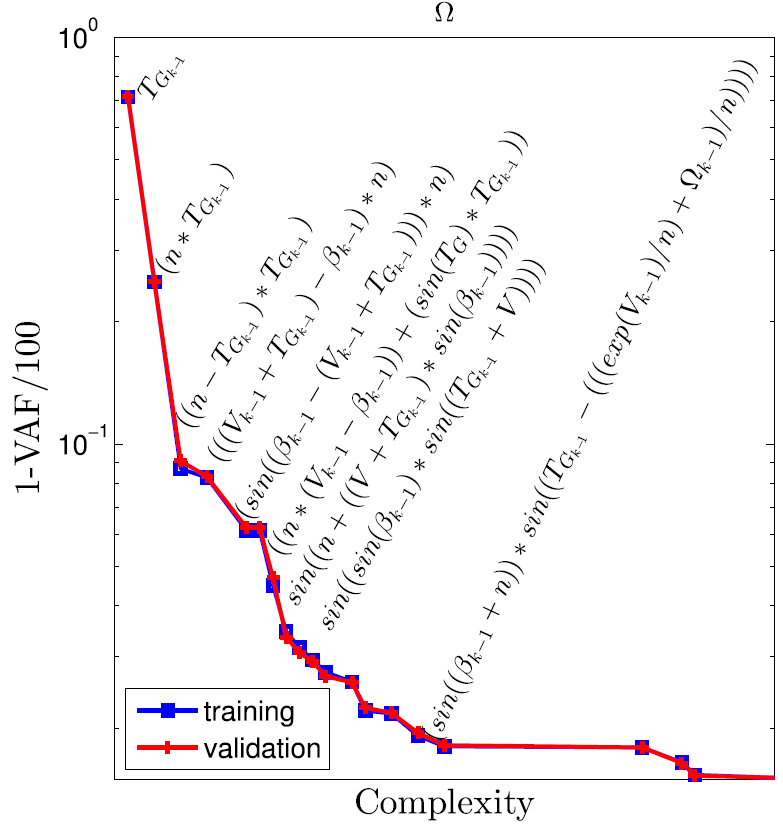}
  \caption{Example Pareto front showing trade-off between solution complexity and variance accounted for (VAF). Reproduced with permission from La Cava \textit{et al.}\cite{LaCava2016} }
  \label{fig:pareto}
\end{figure}

La Cava \textit{et al.}\cite{LaCava2016} applied GPSR to identify nonlinear governing equations of wind turbines. The Pareto front from their paper is reproduced in \autoref{fig:pareto}. The Pareto front illustrates the trade-off between their model complexity as defined by the number and type of operations in the equation and the normalized variance in the prediction error.
La Cava and other authors\cite{LaCava2016a} also tested modifying standard GPSR with
features from epigenetics, such as passive structure, phenotypic plasticity, and 
inheritable gene regulation. 
These researchers demonstrated their modifications improved the performance over 
standard GPSR by finding compact dynamic equations for synthetic data from nonlinear 
ordinary differential equations as well as real-world systems, e.g.\ 
cascaded tanks, a chemical distillation tower, and an industrial wind turbine. 
GPSR has also been applied to testing the efficient market hypothesis\cite{Chen1997}, 
formulating the synchronization control in oscillator networks\cite{Gout2018}, 
identifying the structure of helicopter engine dynamics\cite{Gray1998}, real-time 
runoff forecasting in France\cite{Khu2001} and Singapore\cite{Liong2002}, designing 
circuits\cite{Miller2000}, predicting solar power production\cite{Quade2016}, finding 
dynamical equations for metabolic networks\cite{Schmidt2011} in both cases where a 
starting model was known and from scratch, modelling global temperature 
changes\cite{Stanislawska2012}, and synthesizing second-order coefficient insensitive 
digital filter structures\cite{Uesaka2000}. 

The existing uses of GPSR within chemistry are more extensive than that for materials science. 
We refer the reader to the review by Vyas, Goel, and Tambe\cite{Vyas2015} for 
further details. 
Some key studies with relevance to materials science are summarized here: 
Langdon and Barrett developed a model for oral bioavailability of a small molecule 
given a few hundred data points from expensive experiments\cite{Langdon2005}. 
Their model based on chemical descriptors showed promise for rapid drug screening, 
but had difficulty generalizing to novel molecules. 
Vyas \textit{et al.}\cite{Vyas2014} also discovered structure-property relationships 
for drug absorption using GPSR. 
Here, they demonstrated $R^{2}$ values comparable to those achieved with artificial 
neural networks and support vector regression. 
Barmpalexis \textit{et al.}\cite{Barmpalexis2011} performed a multiobjective 
optimization using GPSR. 
They found a function mapping levels of 4 polymers to three different properties of a 
pharmaceutical release tablet that was more predictive than a shallow neural network. 
Last, Muzny, Huber, and Kazakov built a correlation model for the viscosity of 
hydrogen as a function of temperature and pressure\cite{Muzny2013}. 

\subsection{Opportunities in materials science}

Materials science has many potential areas where GPSR can be applied for 
the same reasons it find use in other disciplines. 
Nonlinear systems are abundant in materials science. 
Changes in materials properties occuring in response to 
structural, composition, and other external perturbations are frequently 
nonlinear in proximity to phase transitions or for large stimuli. 
For instance, changing the concentration of oxygen vacancies in a 
transition metal compound by an atomic percent can alter its 
ionic or electronic conductivity by orders of magnitude\cite{Markov2007,Nakamura}.
The dynamical behavior of materials performance as a function of time is 
also of broad interest and technological importance, e.g.\ corrosion of 
nickel cathodes under different conditions \cite{Daza2000}. 
These are the areas where a dynamical multivariable model would be ideal 
to understand the correlation among the variables and assist 
optimization of materials properties, e.g.\ corrosion resistance.

\revision{
Frequently, materials scientists look for relationships ($f$) 
among multiple variables with the aim to find some closed-form 
expression such as $y=f(X)$, where $y$ is the objective value and $X$ are a set of variables.
These equations are typically expressed in differential form, e.g., the Schr\"{o}dinger equation ($i\hbar\frac{d}{dt}\left| \Psi(t) \right>=\hat{H} \left| \Psi(t) \right>$) or Newton's second law ($\mathbf{F}=m\frac{d\mathbf{v}}{dt}$).
It has been shown that symbolic regression can generate 
ordinary nonlinear partial differential equations 
for nonlinear coupled dynamical systems
\cite{Bongard2007, Maslyaev2019} as well as approximate ordinary differential equations\cite{10.1007/978-3-662-44303-3_3}.
Meanwhile, it is also often of desire to find conservation laws in physical systems. 
The ability to unearth conservation laws with symbolic regression goes 
beyond the aim of materials property predictions
and helps researchers establish insight into the materials systems they study\cite{Schmidt2009,Schmidt2009a}. 
In fact, we do not necessarily need a rigorous expression of natural laws 
in every case; sometimes an approximation with a simple yet effective expression serves well for the research purpose\cite{von1972local}. 
Symbolic regression could potentially balance the 
trade-off between model accuracy and simplicity, 
and might even help scientists discover new equations that redefine our 
understanding of functional materials in the same way those of Hall and  
Petch and Harper and Dorn changed our understanding of the mechanical 
properties of metals or as more recently how Berry phases and topological band theory changed our understanding of electronic structures.}

As we mentioned earlier, 
materials properties and performance are affected by phenomena that involve 
multiple length scales. 
Most theoretical models are formulated to be optimal at a particular length 
scale. 
However, recent emphasis has been placed on multiscale and hiearchical 
modeling in the 
materials science community \cite{multiscale2015,Yadollahi2015,Moore2014}, 
and there is an increased need for effective, descriptive and predicative, 
multiscale models. 
Symbolic regression techniques are potential solutions to this challenge by 
directly searching for the interactions among variables operating and 
passing between multiple spatial and temporal scales. 
Another possible approach is to utilize existing simulation methods within 
each length scale, while using symbolic regression to find the suitable 
coupling interactions between scales, i.e.\ connecting models of different 
scales.
%

Other applications of GPSR in materials and molecular systems are 
in areas where supervised machine learning has already demonstrated 
usefulness in providing new insight or solutions. 
While machine learning has produced many impressive 
results\cite{Zhuo2018, Alberi2019,Ward2017AtomisticReview}, 
it is commonly understood that ML models exhibit a trade-off between 
performance on prediction metrics and the ability to explain the 
predictions of a model due to the complexity of state-of-the-art models 
like deep neural networks or gradient boosted decision trees. 
GPSR offers a middle ground with comparable performance but with the added 
ability to read and directly interpret the output function. 
GPSR is also well-suited to the development of new 
descriptors\cite{Ghiringhelli2015} for materials properties.
By combining features in a manner best suited to fitting data, new features 
are created that can be used as proxies for the property in question. 
This is also a common application of compressed sensing \cite{Ghiringhelli2017,Ouyang}. 
Compressed sensing differs from GPSR in that while GPSR uses GP to 
iteratively evolve a solution, compressed sensing tries to enumerate as 
many combinations of primary features as possible and then use sparsifying 
operators to find a small dimensional subset that correlates with the 
target.  

Materials scientists are not just interested in making predictions. 
They also want to identify the controlling features of a property and 
what role each feature plays; they want to understand which 
degrees-of-freedom to design or optimize to achieve targeted 
properties. 
Given some desired properties as objectives and the relevant variables, 
there are a number of numerical algorithms available for performing 
optimization \cite{vanderplaats2001numerical}, but they typically require 
some a-priori knowledge of the mathematical relationships within the system.
%
The symbolic equations derived from GPSR offer insight into which 
microscopic or macroscopic knobs to turn for the design of desired functionality, such as corrosion resistance in steels\cite{Shimada2002}. 

Some novel ideas include targeting physical variables for 
which we do not know the proper mathematical expressions. 
For instance, the exchange-correlation functional used in density 
functional theory, or the correlation function 
for the viscosity of normal hydrogen \cite{Muzny2013}. 
Contraindicated material property pairs such as 
ferromagnetism and ferroelectricity 
or optical transparency and electrical conductivity 
would be interesting areas to pursue in search for routes to decouple 
or circumvent perceived coexistence incompatibilities. 
In these cases, GPSR could provide candidate representations of these 
terms, where our physical knowledge can be used to filter meaningful 
results. 
One advantage of using symbolic regression is that the balance 
between model accuracy and complexity is tunable with GPSR settings, 
e.g.\ stopping criteria, penalty on individual size, etc. This advantage is particularly evident when designing with constraints or optimizing for performance. The recommendations coming from the learned symbolic equations are more actionable than learned functions only optimized for test set accuracy since they are rigorously made to use fewer terms. 

\subsection{Use cases in materials science}\label{sec:use_case}

\subsubsection{\revision{Discovering the Johnson-Mehl-Avrami-\\
Kolmogorov equation}}
We now show how to use genetic programming-based symbolic regression to 
learn the Johnson-Mehl-Avrami-Kolmogorov (JMAK), hereafter, Avrami equation. %
The Avrami equation quantitatively describes the growth kinetics of 
phases in materials at constant temperature. 
Here, we specifically study the recrystallization process of copper, the 
original experimental data was obtained from Ref.\ \onlinecite{Decker1950}. 
We expect the form of the function to be
\[
  y = 1 - \exp(-kt^n)\,,
\]
where the phase fraction of transformation $y$ is a function of time $t$. 
The coefficients $k$ and $n$ are unknown and change with respect to 
temperature and other environmental conditions.
%

We use GPSR as implemented in \texttt{gplearn} \cite{Stephens2016a} to 
predict the relationship between fraction transformed $y$ and time $t$. 
The hyper-parameters used in GPSR are listed in \autoref{tab:parameters_gp}. %
When the population size is divided by the tournament factor, one obtains 
the number of individuals competing for reproduction each round. 
The parsimony coefficient regularizes the size of individuals by penalizing 
over-sized structures. 
We include  operations of addition, subtraction, multiplication, negation, 
and the natural exponential function into the function set. 
The power function is not included since it easily causes numerical overflow or invalid operations (e.g. power(-1,0.5)).
In general, this can be an issue when evaluating power-law dependent phenomena such as electrical transport equations. Working with log transforms of the original variables may be a more stable approach. 
Crossover operations dominate the genetic operations with a 70\% probability
to be applied; the other 30\% chance corresponds to mutation operations 
(e.g.\ point mutation, subtree mutation, etc.). 
Additional details concerning the usage of \texttt{gplearn} are given 
in its documentation.

\begin{table}[t]
\begin{ruledtabular}
\caption{List of hyper-parameters used in GPSR to learn the 
Avrami equation. Grid search method is used to find the optimal hyper-parameters from the top three parameter sets.
\label{tab:parameters_gp}
}\centering
\begin{tabular}{ll}
\sffamily Parameter & \sffamily Values  \\
\hline
population size & \{2000, 5000\} \\
tournament factor & \{100, 500\} \\
parsimony coefficient & \{0.001, 0.005\} \\
max generation & 20 \\
constant range & (-1, 1) \\
function set & \{add, sub, mul, neg, exp\} \\
crossover probability & 0.7 
\end{tabular}
\end{ruledtabular}
\end{table}

Data preprocessing is an essential step in machine learning before feeding 
data into the solver. 
Conventional preprocessing methods include shifting data to be 
zero-centered, and scaling data to unit standard deviation. 
However, the conventional preprocessing methods are not ideal choices in 
our case. 
Zero-centered shifting is not applicable to either the phase fraction 
transformed ($y$) or time ($t$) since we want to obtain the exact function 
form of the Avrami equation. 
Furthermore, scaling the time frame would only change the constant $k$ in 
the Avrami equation. 
Here, we scale time from 0 to 10 for all data sets so that the 
constant $t$ lies in the range [-1, 1].
The $y$ values remain unchanged as the experimental data 
range over [0, 1].

\begin{figure}[t]
  \centering
 \includegraphics[width=0.99\linewidth]{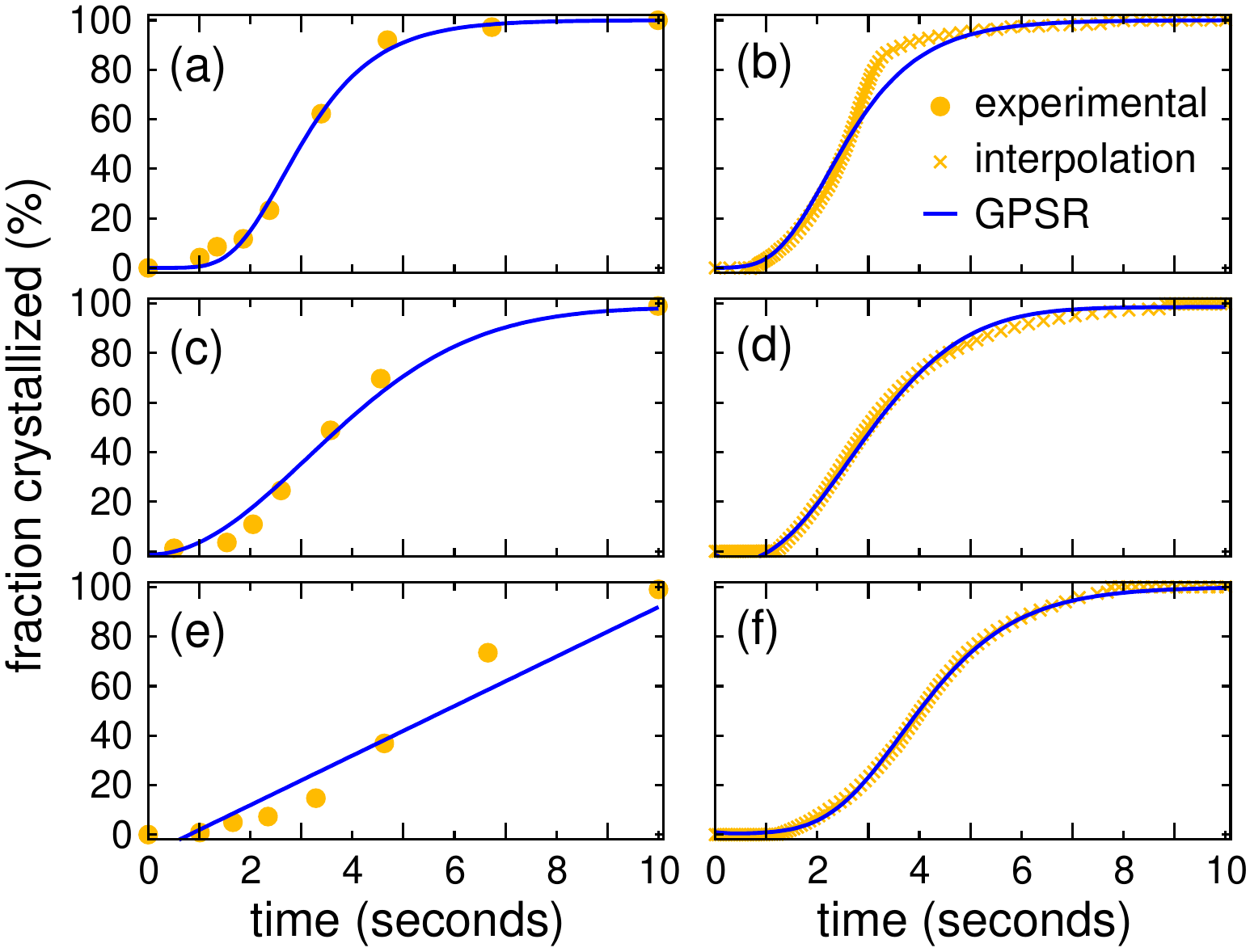}
  \caption{GPSR prediction and performance with different data sets. Left panels correspond to the direct experimental data while right panels use interpolated experimental data 
  (a) 135$^{\circ}$C experimental data, 
  (b) 135$^{\circ}$C extrapolated data, 
  (c) 113$^{\circ}$C experimental data, 
  (d) 113$^{\circ}$C extrapolated data, 
  (e) 102$^{\circ}$C experimental data, and 
  (f) 102$^{\circ}$C extrapolated data.
  }
  \label{fig:avrami_results}
\end{figure}

\begin{table*}
\begin{ruledtabular}
\caption{Functions predicted by GPSR and numerical fitting results for the Avarmi equation describing copper recrystallization.\label{tab:gpsr_results}}
\centering
\begin{tabular}{llll}
\sffamily Dataset & \sffamily Numerical fitting result & \sffamily GPSR result & \sffamily GPSR (transformed $y$) result \\
\hline
ideal & $y = 1-\exp(-0.6t^2)$ & $y = 0.994-\exp(-0.58t^2)$ & $1-y = \exp(-0.58 t^2)$\\
\hline
135$^{\circ}$C experimental & $y = 1-\exp(-0.019t^{3.2})$ & $y = \exp[-\exp(-t+2.64)]$ & $1-y = \exp(-0.024 t^3)$ \\
135$^{\circ}$C interpolated & $y = 1-\exp(-0.17t^{3.6})$  & $y = \exp[-\exp(-t+2.18)]$ & $1-y = \exp[-0.065 t^2 (t-0.851)]$\\
\hline
113$^{\circ}$C experimental & $y = 1-\exp(-0.018t^{2.8})$ & $y = 0.986-\exp(-0.051t^2)$ & $1-y = \exp(-0.014 t^3)$ \\
113$^{\circ}$C interpolated & $y = 1-\exp(-0.042t^{2.5})$ & $y = 0.986-\exp[-0.108 \times t \times (t-0.934)]$ & $1-y = \exp[-0.097 t (t-0.764)]$\\
\hline
102$^{\circ}$C experimental & $y = 1-\exp(-0.0054t^{2.9})$ & $y = 0.1t - 0.081$ & $1-y = \exp(-0.0045 t^3)$\\
102$^{\circ}$C interpolated & $y = 1-\exp(-0.01t^{3.1})$ & $y = \exp[-(1.11+2t)\times\exp(-t+1.42)]$ & $1-y = \exp(-0.01 t^3)$\\
\end{tabular}
\end{ruledtabular}
\end{table*}

We take the experimental copper recrystallization data at temperatures  
135$^{\circ}$C, 113$^{\circ}$C, and 102$^{\circ}$C as input and scale the 
time variable before performing regression. 
Since the experimental data only contains several points at each
temperature (less than 10), we also take data points from 
interpolated lines and perform symbolic regression on the interpolated
data for comparison. 
An ideal dataset is also generated, where $y$ values are directly 
calculated as $1-\exp(-0.6 t^2)$. 
The optimal $k$ and $n$ values in each data set are 
obtained from numerical fitting using \texttt{Scipy}, 
given the form of the Avrami function.
Ideally GPSR should be able to recover the correct function form as well as 
$k$ and $n$ constants for all data sets. 
The best individual after 20 generations of evolution is collected for 
each hyper-parameter setting. 
Finally we manually pick the optimal individual with closest function form 
and constants to the Avrami equation within each data set. 
Our parameter fitting and GPSR evolution results are shown in 
\autoref{tab:gpsr_results} and \autoref{fig:avrami_results}.

\texttt{Gplearn} successfully recovers the relationship between time ($t$) 
and fraction transformed ($y$) in most cases. 
With the ideal input data, \texttt{gplearn} evolves almost a 
perfect form of the Avarmi function as well as the constants 
(see the first row of \autoref{tab:gpsr_results}). 
The performance on the raw experimental and interpolated data are 
generally worse than the ideal case but still reasonable. 
In most cases the exponential function form and polynomial function of 
$t$ are both recovered. 
One source of error is the lack of the power function in function set, 
which was intentionally omitted as non-integer powers of variables cannot 
be correctly represented here.
Rather, polynomial functions are used as an approximation. 
With the limited choice of mathematical functions, GP produces results that 
are very close semantically but exhibit different syntax.
\revision{
We see in multiple cases GPSR produces functions of the 
$\exp[-\exp(\Theta)]$ form,  where the expected function form is $1-\exp(\Theta)$. 
Here we introduce a mathematical trick to show their 
equivalence. 
The functions $e^x$ and $1+x$ are called 
equivalent infinitesimals because
\[
\lim_{x \to 0} e^x \sim 1+x\,.
\]
In our case, the exponent $-\exp(\Theta)$ quickly reaches zero when $\Theta$ (ideally having the form $-kt^n$) 
becomes more negative; therefore, the equivalent infinitesimal relationship 
holds and the predicted form of the function is equivalent to the expected 
one appearing in the Avrami expression. 
Based on this example, we encourage professional materials researchers, who 
are novice data scientists, to perform careful analysis of GPSR results when 
making the final interpretation of the obtained model.  
}

In real-world applications where the actual function form in unknown, 
the exact syntax of GPSR results may not matter. 
There are potentially many cases where GPSR would produce semantically 
similar/equivalent functions that vary in their syntax, i.e.\ different 
function form. 
Considering the trade-off between model accuracy and complexity, 
in some cases it might be a virtue to find a simple approximated solution 
instead of using rigorous but complex relationships among multiple variables.

The GPSR result from the 102$^{\circ}$C experimental data set (\autoref{fig:avrami_results}e), 
turns out to be a linear relationship and deviates significantly from the original data. 
The poor result may be caused by the relatively small slope of the original data and the insufficient number of available data points. 
Taking the model performance and complexity into consideration, we find 
that the linear relationship survives due to its simple form. 
With interpolated data as input, functions evolved from GPSR agree better with the Avrami equation, as shown in \autoref{fig:avrami_results}f.
\revision{Therefore, having more training data could improve the performance of GPSR, which also applies to other data-driven methods.}
Interestingly, the performance of GPSR can also be improved by transforming (or 
simplifying) the mathematical expression. 
We compare results of directly training $y \sim t$ relationships with $(1-y) \sim t$, 
where in the latter case the target value $1-y$ is the percentage of copper 
untransformed. 
The results are shown in the last column of \autoref{tab:gpsr_results}:
The transformed functions show improved performance since we have already performed 
the subtraction function for the model. 
%
GPSR not only successfully recovers the exponential form of the equation, but also finds constants closer to the numerical fitting values.
%
%
%

\subsubsection{\revision{Learning Landau free energy expansion}}

Next, we present a slightly more complicated case with two variables. The model we studied is the Landau free energy expansion for the cubic-to-rhombohedral structural phase transition in perovskite LaNiO$_3$, where
the free energy $G$ of the system is expanded in powers of an order parameter $\theta$ as 
\begin{equation} \label{eq:landau_dft}
    G(\theta, T) = G_0(T) + \kappa (T - T_C) \theta^2 + \lambda \theta^4 \ ,
\end{equation}
where $\kappa$ and $\lambda$ are temperature-independent coefficients and $\theta$ is the angle of rotation about the [111] direction. This rotation angle of the corner-connected NiO$_6$ is the order parameter for the displacive transition. The parameters we used are obtained from \textit{ab initio} DFT simulations\cite{gou2011lattice}, where $\kappa = 1.696\,\mathrm{meV}\,10^{-3}\,\mathrm{K}/(^\circ)^2$, $\lambda = 0.0171\,\mathrm{meV}/(^\circ)^4$, and $T_C$ is estimated to be $2.057\ (10^3\,\mathrm{K})$. We use $10^3\,\mathrm{K}$ as the unit for temperature so that the constants are brought into a smaller range.
%

In this case, we set the temperature $T$ and order parameter $\theta$ as the input variables, and the free energy change $G(\theta, T) - G_0(T)$ as the output. We uniformly sampled 11 temperature points between [0, 1] ($10^3\,\mathrm{K}$), and 100 order parameter points within range [-20$^\circ$, 20$^\circ$]. The corresponding value for the change in free energy is calculated from \autoref{eq:landau_dft} using the parameters reported in the literature.
A population size of 10,000 is used, with tournament size 25, parsimony coefficient 
0.02, and constant range [-2.0, 2.0]. In order to simplify the problem, we only consider addition, subtraction, and multiplication operations in our search. Other settings are the same to those in the Avrami case.

\begin{figure}[b]
  \centering
 \includegraphics[width=0.99\linewidth]{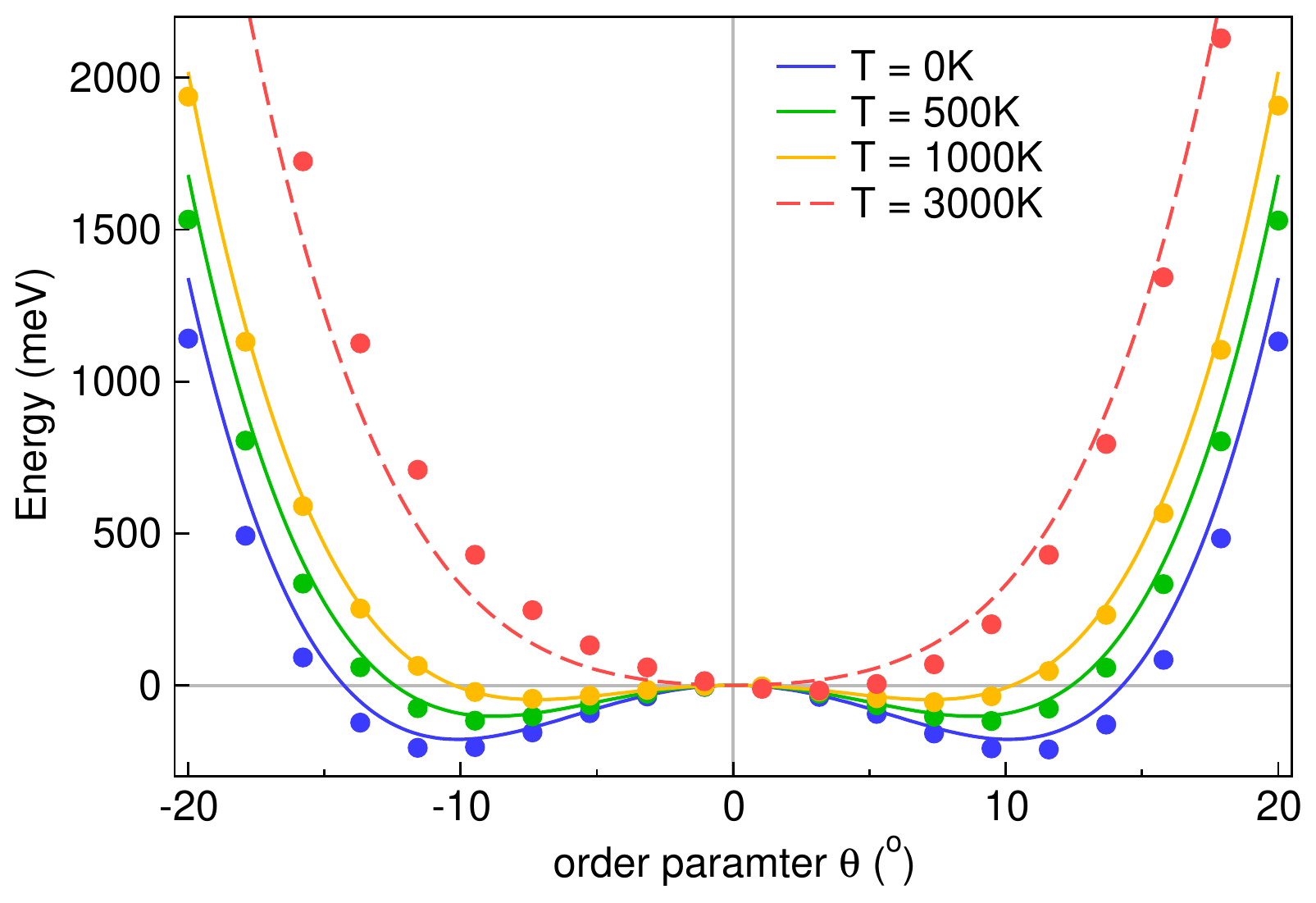}
  \caption{Landau free energy $G$ for perovksite LaNiO$_3$ as a function of the order parameter $\theta$ at different temperatures. Both solid and dashed lines are calculated using \autoref{eq:landau_dft} with coefficients reported in Ref.\ \onlinecite{gou2011lattice}. Only solid lines are used during the training. The filled symbols correspond to GPSR predicted results using \autoref{eq:landau_gpsr}.
  }
  \label{fig:landau_results}
\end{figure}

The best individual after 15 generations of evolution has the function form
\begin{equation}\label{eq:landau_gpsr}
    G(\theta, T) = G_0(T) + 1.983(T - T_C^{'}) \theta^2 + 0.0165 \theta^4 + \xi\,,
\end{equation}
where $T_C^{'} = 1.894 + 8.32\times 10^{-3}T^2$, and $\xi = (- 1.72T^2 + 1.214T - 0.24)\theta - 1.334$.
The GPSR-learned coefficients are quite close to the reported values, especially for the quartic term. 
This is probably because the penalty for a larger deviation in the leading (quartic) term is much higher than others. For the quadratic term of $\theta$, GPSR successfully captured the coupling term $T \theta^2$ and its coefficient, but also an unexpected biquadratic $T^2 \theta^2$. It should not have a strong impact on the function owing to its small coefficient ($8.32\times 10^{-3}$). 

%
The Landau free energy with respect to the order parameter is plotted in \autoref{fig:landau_results}.
We find that GPSR results (filled symbols) agree well with the DFT-derived Landau free energy function not only within the training region, i.e., with $T = 0, 500$ and $1,000\,\mathrm{K}$, but also reasonably well beyond it.
The dashed red line in \autoref{fig:landau_results} with $T = 3,000\,\mathrm{K}$ is not included in training the GPSR,
yet the model reproduces both the shape and the correct global minimum position very well.
However, the predicted function contains other coupling terms not present in \autoref{eq:landau_dft}.
These extra terms in $\xi$ destroy the symmetry of the function, i.e., the free energy expansion is an even function by symmetry. This would become an issue especially when $T$ is large, as we can see a minor shift to positive $\theta$ values occurs for the red filled symbols in \autoref{fig:landau_results}. Again, the model may need more data to learn the correct even function form. In general, GPSR does an excellent job in learning the relationship between temperature and the order parameter without any knowledge of the physical system. Our results here also reveal the potential to perform effective feature selection using GPSR, where the insignificant variables (features) could be approximated or ignored in post-processing (e.g. terms with negligibly small coefficients).

The purpose of these two use cases for learning the 
Avrami equation and Landau free energy expansion with GPSR is 
to show the potential of its application in materials science problems. 
Although these are relatively simple examples, the understanding and approaches 
applied could be generalized and utilized to solve real-world challenges. 
Apart from both function analysis and data pre-processing mentioned previously, 
hyper-parameter tuning is also an essential step to obtain the optimal solution. 
A grid-search scheme to search the hyper-parameter space 
(e.g., population size, number of generations, regularization, etc.) 
is recommended since the optimal settings differ from case to case. 
The grid-searching process can be exhausting, but it might also be rewarding.
Comparing the results from different hyper-parameter settings could provide insight into the functional form of the optimal solution, 
especially if components of a particular function appears multiple times in the 
solution set.

The real challenge, however, is that very often materials science problems cannot be represented using regular functions (analytic and single-valued). A simple example would be to understand how different chemical compositions affect materials properties\cite{yu2016electronic}. This problem originates from the inability to differentiate in the chemical space and is out of the scope of this prospective. It remains an open question in the materials research community.

\section{Summary} \label{sec:4}

Symbolic regression has shown competitive performance to other machine learning-based regression models in various research domains. While there are some shortcomings of the current state-of-the-art GPSR, e.g.\ high computational cost, non-deterministic optimization, there are numerous active research efforts focusing on improving the performance of symbolic regression to expand its use in real-world applications. The ability of symbolic regression to distill natural laws from data sets with high-dimensional parameter space makes it an ideal technique for materials science research, since  
these researchers typically face sparse data sets with multiple variables. Freed from having a fixed form of equations, symbolic regression can potentially reveal the significant interactions among physical variables.  We recommend more materials scientists utilize symbolic regression techniques in their own research domain.

\begin{acknowledgments}
Y.W.\ acknowledges partial support from the 
Predictive Science and Engineering Design (PS\&ED) program at Northwestern University. %
All authors acknowledge support from the National Science Foundation (NSF) through 
the Designing Materials to Revolutionize and Engineer our Future (DMREF) program 
under award number DMR-1729303.
\end{acknowledgments}

\bibliography{reference}

\end{document}